\documentclass[twocolumn,twocolappendix]{aastex701}

\usepackage{booktabs}
\usepackage{soul}
\usepackage{threeparttable}

\usepackage{amsmath}
\usepackage{graphicx}
\usepackage{dcolumn}
\usepackage{bm}
\usepackage{hyperref}
\usepackage{color}
\usepackage{aas_macros}
\usepackage{mathtools}

\newcommand{\lamC}{\ensuremath{\lambdabar_\mathrm{C}}}
\newcommand{\ud}{\mathrm{d}} 
\renewcommand{\vec}[1]{ \bm{#1}} 
\newcommand{\nvec}[1]{ \hat{\bm{#1}}}

\def\bfnabla{\nabla}

\newcommand{\unitspace}{\,}

\newcommand{\g}{\ensuremath{\unitspace \mathrm{g}}}

\shortauthors{Salmi \& N\"attil\"a }

\begin{document}

\title{Pair Discharges and Radio Emission from Millisecond-Pulsar and White-Dwarf Magnetospheres}

\author[0000-0001-6356-125X ]{Tuomo~Salmi}
\affil{Department of Physics, University of Helsinki, P.O. Box 64, FI-00014 University of Helsinki, Finland}
\email{tuomo.salmi@helsinki.fi}

\author[0000-0002-3226-4575] {Joonas Nättilä}
\affil{Department of Physics, University of Helsinki, P.O. Box 64, FI-00014 University of Helsinki, Finland}
\email{joonas.nattila@helsinki.fi}

\begin{abstract}
Coherent radio emission is observed from compact astrophysical objects with relatively weak magnetic fields, including millisecond pulsars and white dwarfs, the latter being proposed as possible sources of long-period radio transients.
In such environments, the standard pair discharge mechanism---driven by curvature radiation and one-photon pair production---can fail because the low magnetic field strength suppresses photon conversion. 
We analyze a discharge mechanism that operates efficiently in weak-field magnetospheres, including two-vertex quantum electrodynamic processes: inverse Compton up-scattering of background photons followed by either two-photon or one-photon pair creation depending on the seed photon temperature. 
Using first-principles radiative particle-in-cell simulations incorporating exact QED cross sections, we demonstrate that these mechanisms robustly generate pair cascades in millisecond-pulsar magnetospheres and in hot white-dwarf environments.
The resulting system exhibits limit-cycle behavior and generates electromagnetic field fluctuations capable of producing coherent radio emission, providing a natural explanation for radio activity in low-field compact objects.
\end{abstract}


\section{Introduction}

Coherent radio emission is observed from a wide range of compact astrophysical objects, including neutron stars (NS; \citealt{philippov2022}) and white dwarfs (WDs; \citealt{zhang2005,hurleywalker2022,deruiter2025}).
Pair discharges in the radio pulsar---NSs with relatively strong magnetic $\vec{B}$ fields---magnetospheres have been shown to generate strong electromagnetic fluctuations that can drive such coherent emission \citep{philippov2020,cruz2021,benacek2024,chernoglazov2024,benacek2025,ye2025}.
These discharges occur in \emph{gaps}--—regions of unscreened electric field $\vec{E}_\parallel$ parallel to $\vec{B}$ where particles are accelerated to ultra-relativistic energies. 
The gap field strength is typically a fraction of the rotation-induced electric field $E_\mathrm{rot} \sim \Omega R B/ c$, where $\Omega$ is the stellar angular frequency and $R$ its radius.

In strongly magnetized systems, such as many pulsars, the accelerated particles emit curvature or synchrotron photons that undergo one-photon pair creation, $\gamma + [B] \rightarrow e^+ + e^-$, producing electron–positron cascades (here, $\gamma$ depicts photons, $e^\pm$ pairs, and $[B]$ is a catalyst for the process). 
The efficiency of this process depends critically on the local magnetic field strength relative to the quantum electrodynamic (QED) critical field $B_Q \equiv m_e^2 c^3 / (e \hbar) \simeq 4.4 \times 10^{13} \,\mathrm{G}$. 
For ordinary radio pulsars with $B \sim 10^{12} \,\mathrm{G} \sim 0.02 B_Q$, this process proceeds efficiently.
The resulting gaps are intermittently screened and exhibit limit-cycle behavior, producing quasi-periodic plasma and electromagnetic fluctuations that have been proposed as a source of pulsar radio variability \citep{philippov2020,tolman2022,okawa2024}.

\begin{figure*}[t!]
\centering
\includegraphics[clip, trim=0.0cm 0.0cm 0.0cm 0.0cm, height=0.5\textwidth]{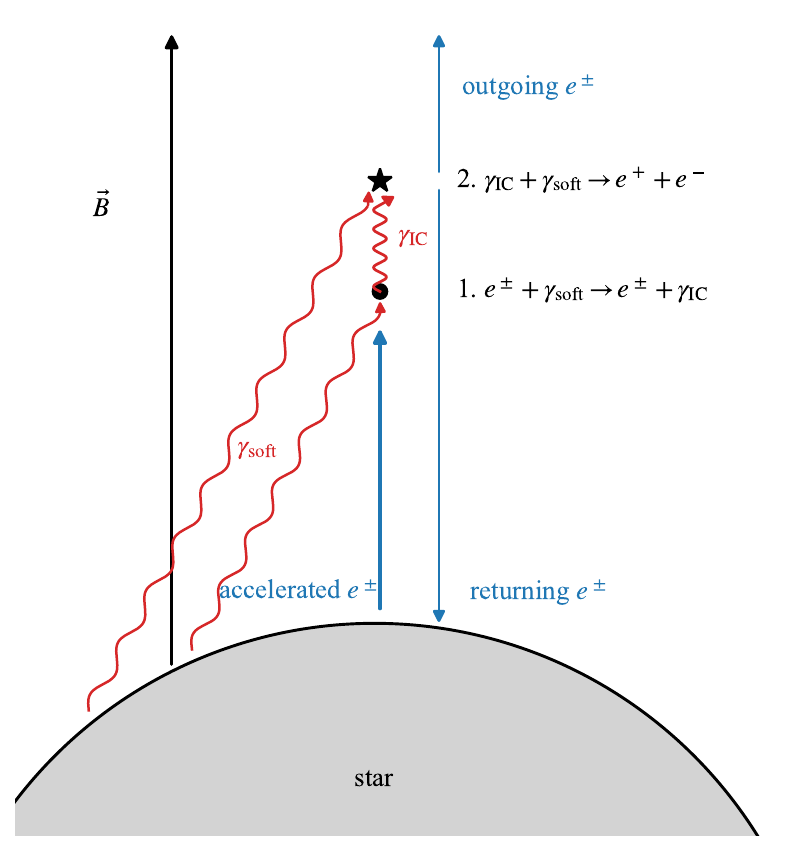}
\includegraphics[clip, trim=0.0cm 0.0cm 0.0cm 0.0cm, height=0.5\textwidth]{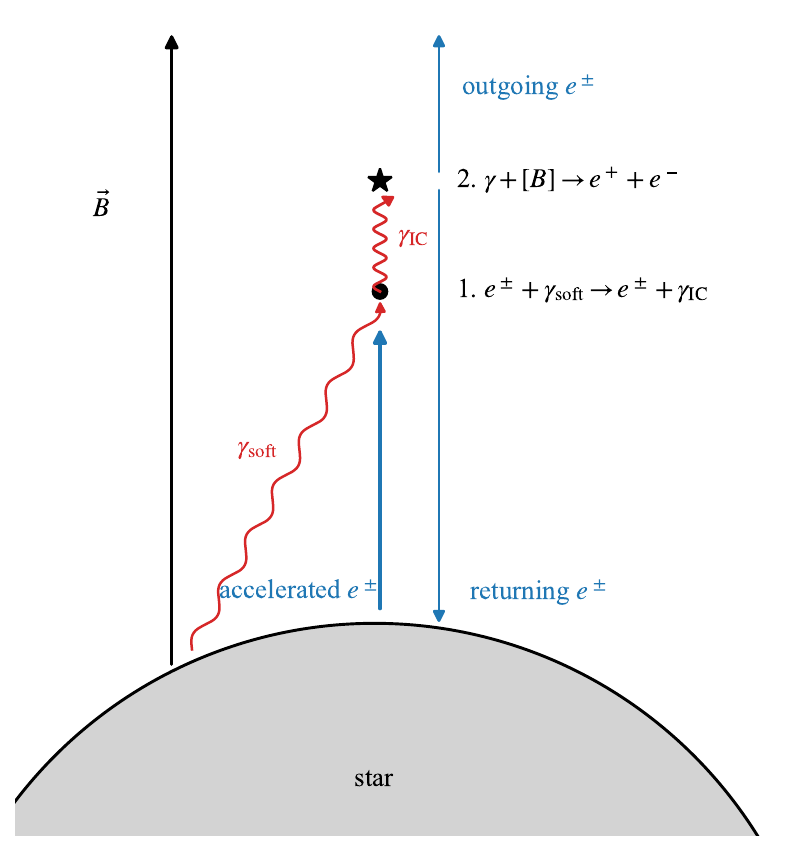}
\caption{\label{fig:gap_schematic} 
Schematic illustration of the Compton-initiated pair-discharge channels above a stellar polar cap. 
Particles, shown by blue straight arrows, are accelerated along the open magnetic field and inverse-Compton scatter soft photons, shown by red wavy arrows, producing high-energy photons $\gamma_{\rm IC}$. 
These photons create pairs either through two-photon pair production, $\gamma_{\rm IC}+\gamma_{\rm soft}\rightarrow e^+ + e^-$, shown on the left, or through magnetic one-photon conversion, $\gamma_{\rm IC}+[B]\rightarrow e^+ + e^-$, shown on the right. 
Some of the produced particles escape outward, while others return to the stellar surface and heat the polar cap.
}
\end{figure*}

However, many compact objects possess much weaker magnetic fields of $B \sim 10^8 \ll B_Q$, including millisecond pulsars (MSPs), WDs, and black holes; nevertheless, coherent radio emission is still observed in such systems.
The origin of this emission in low-field systems remains uncertain.
Here we analyze an alternative pair-discharge channel that proceeds via inverse Compton up-scattering of external background photons to high energies, followed by either two-photon pair creation, $\gamma_{\rm IC}+\gamma_{\rm soft}\rightarrow e^+ + e^-$, or magnetic one-photon pair creation, $\gamma_{\rm IC}+[B]\rightarrow e^+ + e^-$ (see Figure~\ref{fig:gap_schematic}). 
While the two-photon mechanism has previously been explored in simulations of supermassive black hole magnetospheres \citep{levinson2018,yuan2025}, we show that it is more general and can operate in hot MSPs, WDs, and other low-field systems, potentially also explaining some of the long-period radio transients (LPTs, \citealt{rea2026}).
Especially the first variant with Compton-initiated one-photon conversion can be the dominant channel in cooler MSPs.
The resulting discharge-produced pairs are also responsible for heating the MSP atmosphere \citep{zhang2000,gonzalez2010,baubock2019,salmi2020} generating the thermal X-ray emission measured by Neutron Star Interior Composition ExploreR (NICER) and other missions to constrain NS radii.
Because the density of back-flowing pairs in our scenario is relatively low, the model avoids the longstanding polar-cap overheating problem of pulsars, in which theoretical heating rates typically exceed observational constraints \citep[e.g.,][]{beskin2018}.

\section{Simulation Setup}\label{sect:setup}

We model the gap dynamics using first-principles radiative particle-in-cell (PIC) simulations with the \textsc{runko} framework \citep{runko}. 
The system self-consistently evolves electrons ($e^-$), positrons ($e^+$), protons ($p$), and photons ($\gamma$), including Compton scattering and pair creation via exact QED differential cross sections sampled with an adaptive Monte Carlo scheme \citep{nattila2024}.

The computational domain is a one-dimensional region above the magnetic pole, extending over $L_\mathrm{box} = 1.2 H_\mathrm{gap}$, where $H_\mathrm{gap} \lesssim R$; particles are confined to move along this direction, but the full three-component electromagnetic fields are evolved and photon momenta are treated as three-dimensional vectors.
We adopt a uniform background magnetic field $\vec{B}(x) = B_0 \nvec{x}$. 
While realistic NS and WD magnetospheres may exhibit dipolar or multipolar structure, this approximation isolates the essential gap physics.

The electromagnetic fields evolve according to Maxwell’s equations in a co-rotating frame \citep{Nattila2026}:
\begin{align}
\partial_x E_x &\approx 4\pi (\eta - \eta_\mathrm{co}), \label{eq:maxwell1}\\
\partial_x B_x &= 0, \label{eq:maxwell2}\\
\partial_t \vec{B} &= - c \bfnabla \times \vec{E}, \label{eq:maxwell3}\\\
\partial_t \vec{E} &\approx -4\pi (\vec{j} - \vec{j}_m) \,. \label{eq:maxwell4}
\end{align}
The frame moves with velocity $\vec{v} = \beta_\mathrm{rot} c \,\nvec{y}$, representing stellar rotation at cylindrical radius $R_\mathrm{pc}$, such that $\beta_\mathrm{rot} \approx \Omega R_\mathrm{pc}/c$.
For typical parameters (assuming $R_\mathrm{pc}$ is the radius of the polar cap region), $\beta_\mathrm{rot} \sim 3\times10^{-2}$ (MSPs) and $\sim 3\times10^{-6}$ (WDs).
Rotation induces the Goldreich–Julian co-rotation charge density $\eta_\mathrm{co} \approx -\beta_\mathrm{rot} B_0 / (4\pi R_\mathrm{pc})$ \citep{goldreich1969}, while large-scale magnetic twist is represented by an external current $\vec{j}_m = c \bfnabla \times \vec{B}/4\pi$.

A steady, gap-free state requires $\eta = \eta_\mathrm{co}$ and $\vec{j} = \vec{j}_m$. 
Deviations are expected to arise for $\alpha \equiv j_m/(\eta_\mathrm{co} c) > 1$ (or $\alpha < 0$), leading to unscreened electric fields under free particle flow \citep{mestel1985,belmont2008,beloborodov2008,timokhin2013}. 
We adopt $\alpha = 1.5$ as our fiducial value and set the initial electric field to increase linearly from $\vec{E}(0)=0$ to $\vec{E}(H_\mathrm{gap}) = E_\mathrm{rot}\nvec{x}$, where $E_\mathrm{rot} \approx \beta_\mathrm{rot} B_0$, corresponding to a voltage drop $|\Delta V| = \tfrac{1}{2} E_\mathrm{rot} H_\mathrm{gap}$. 
The initial charge density is set to satisfy Equation~\eqref{eq:maxwell1}, and pair creation is restricted to $0 < x < H_\mathrm{gap}$ to model finite gap height.

Unlike in most pulsar gap models, the stellar surface acts in our case as both a particle and photon reservoir.
In the fiducial setup, we inject thermal electrons and protons with dimensionless temperature $\theta_\mathrm{p} = k_\mathrm{B}T_p/(m_e c^2)=0.3$ along $\vec{B}$ \citep{arons1979} (simulations are found to be insensitive for this parameter when $\theta_\mathrm{p} < 1)$. 
The surface additionally emits isotropic thermal photons with temperature $\theta_\mathrm{BB} = k_\mathrm{B}T_\mathrm{BB}/(m_e c^2)$.
For computational efficiency, photons are injected at a reduced rate $\dot{n}_\mathrm{inj}$, while QED interaction rates are rescaled so that the Compton and pair-production mean free paths match those expected in astrophysical systems (see Appendices \ref{app:estimates} and \ref{app:supplementary}). 
In our baseline model, we neglect one-photon pair production and adopt parameters representative of hot MSPs ($k_\mathrm{B}T_\mathrm{BB}\sim 0.3$ keV) and WDs ($\sim 0.05$ keV). 
The adopted WD temperature should be interpreted as characterizing a very hot, localized (possibly non-thermal) polar-cap photon field rather than the global surface of a typical WD.

The simulation uses $L_\mathrm{box}/\Delta x = 20480$ cells, decomposed into $1024$ subdomains. 
The co-rotation density $n_\mathrm{co} = |\eta_\mathrm{co}/e|$ is set to correspond to 16 particles per cell per species, resolving the skin depth $c/\omega_{\mathrm{p,co}}$ with $\sim 17$ cells, where $\omega_{\mathrm{p,co}} = (4\pi n_\mathrm{co} e^2/m_e)^{1/2}$. 
To reduce numerical noise, the current is smoothed with 8 passes of a binomial filter, effectively increasing the particle statistics. 
Additional numerical details are provided in Appendix \ref{app:supplementary}.

\begin{figure*}[t!]

\centering
\hspace*{-1.45cm}
\includegraphics[clip, trim=0.0cm 0.0cm 0.0cm 0.0cm, height=1.0\textwidth]{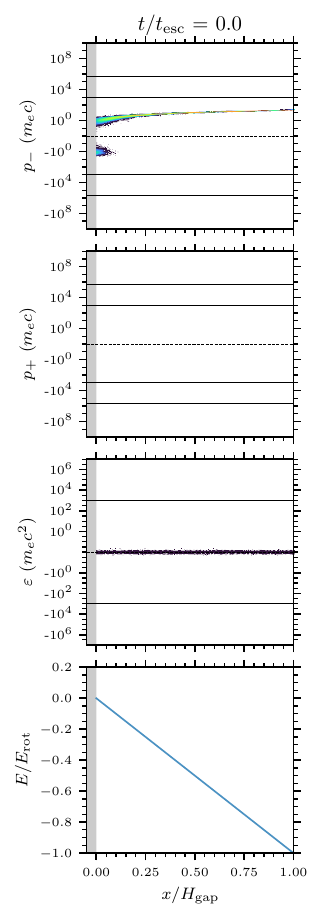}
\hspace*{-0.66cm}
\includegraphics[clip, trim=1.30cm 0.0cm 0.0cm 0.0cm, height=1.0\textwidth]{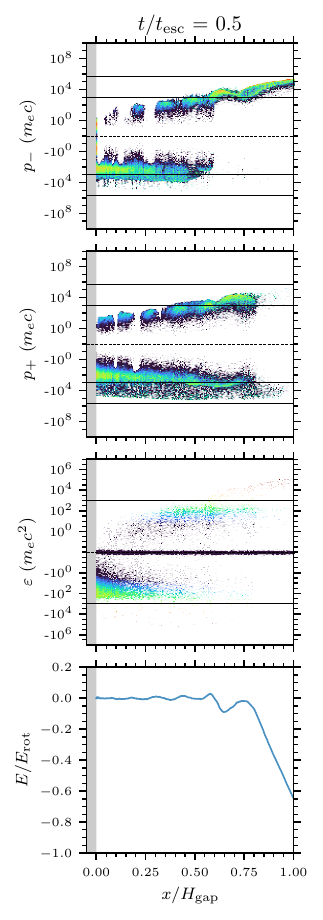}
\hspace*{-0.66cm}
\includegraphics[clip, trim=1.30cm 0.0cm 0.0cm 0.0cm, height=1.0\textwidth]{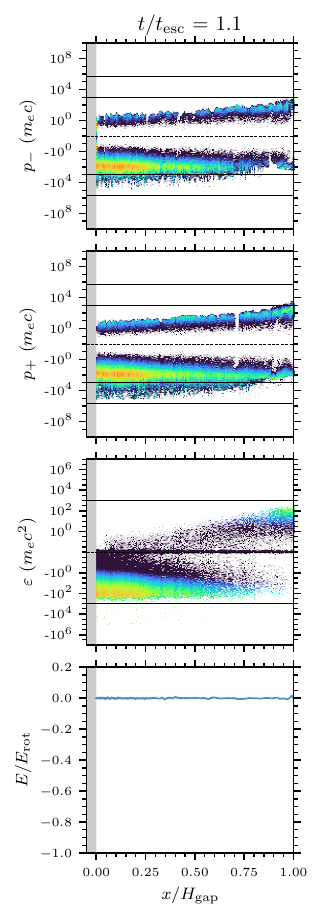}
\hspace*{-0.66cm}
\includegraphics[clip, trim=1.30cm 0.0cm 0.0cm 0.0cm, height=1.0\textwidth]{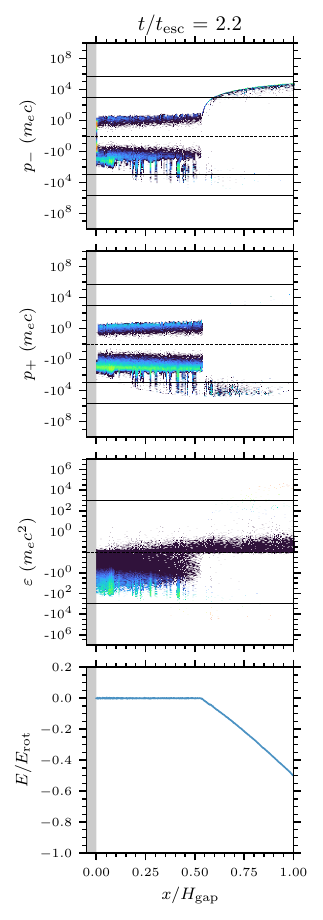}
\caption{\label{fig:gap_sim} 
Simulation snapshots of the discharge cycle.
The first row shows the electron distribution in the $x–p$ phase space, and the second row shows the corresponding positron distribution. 
Colors indicate the local particle density (proportional to $\mathrm{d} n(p)/\mathrm{d} \log p$).
The third row shows the photon energy spectra, with colors indicating the local photon density (proportional to $\varepsilon \,\mathrm{d} n_{\varepsilon}(\varepsilon)/\mathrm{d} \log \varepsilon$).
The values range from low to high as black to blue to green to red.
In all three rows, the sign of the vertical axes denotes the direction of particle motion or photon propagation along the x-axis; the logarithm is taken of the corresponding absolute momentum or photon energy.
Vertical thin lines show the $\gamma_\mathrm{gap}$, $\gamma_\mathrm{thr} = \varepsilon_\mathrm{thr}= 10^{3}$ values (matching the simulated MSP scenario), and the dashed lines the $p = 0$ value.
The bottom row shows the gap electric field $E_x$ in units of $E_\mathrm{rot}$. 
}
\end{figure*}

\section{Radiative Particle Dynamics}

In the simulations, we observe that electrons and positrons are rapidly accelerated within the gap. 
In the ultra-relativistic limit, their energy evolution is governed by
\begin{equation}
    \frac{\ud \gamma}{\ud t} \simeq \frac{q E_x}{m_e c} - \frac{P_\mathrm{rad}}{m_e c^2} \, ,
\end{equation}
where $\gamma$ is the Lorentz factor, $q = -e$ for electrons and $q = +e$ for positrons, and $P_\mathrm{rad}$ the radiative loss power.
In the absence of radiative losses, the available potential drop $\Delta V$ implies a maximum Lorentz factor $\gamma_{\mathrm{gap}} \sim 10^{8}$ for MSPs and $\sim 10^{6}$ for WDs (see Appendix \ref{app:estimates}). 
Our scaled simulations correspond to $\gamma_{\mathrm{gap}} \sim 10^{6}$.

For both MSP and WD conditions, radiative losses do not strongly limit particle acceleration. 
Curvature radiation would yield an equilibrium Lorentz factor $\gamma_{\mathrm{rad}} \sim 10^{7}$, but we neglect this channel for simplicity. 
Inverse Compton (IC) losses are suppressed in the Klein–Nishina regime and therefore do not significantly affect the particle energetics. 
However, IC scattering remains crucial as the primary mechanism for generating high-energy photons, and is included in the simulations using the exact anisotropic Compton formalism. 

Thermal photons with characteristic energy $\varepsilon_\mathrm{BB} \approx 2.7\,\theta_\mathrm{BB}$ are up-scattered to \citep{rybicki1979,aharonian2004}
\begin{equation}
\varepsilon_{\mathrm{IC}} \approx \frac{4\gamma^{2}\varepsilon_{\mathrm{BB}}}{1 + 4 \gamma \varepsilon_{\mathrm{BB}}}
\end{equation}
for head-on collisions. 
In our regime, a single scattering typically extracts a substantial fraction of the particle energy, boosting photons to energies near or above the peak of the two-photon pair-production cross section, $\varepsilon_{\gamma\gamma} \approx 3.7/\varepsilon_\mathrm{BB}$ \citep{coppi1990}.
With our baseline model the mean-free path for the two-photon process is $l_\mathrm{mfp}/H_\mathrm{gap} \sim 10^{-1}$ when $\gamma \sim 0.01 \gamma_\mathrm{gap}$.
For MSP parameters, a single scattering of a maximally accelerated particle can additionally trigger one-photon pair creation within a distance shorter than $H_\mathrm{gap}$ (see Appendix \ref{app:estimates}).
We treat this channel separately below.
Generally, we find that multiple scatterings are not required to initiate pair cascades.

\section{Pair Cascades}

Pair creation can trigger a cascade with copious $e^\pm$ production, as newly created pairs are rapidly accelerated toward $\gamma_\mathrm{gap}$ and, in turn, up-scatter additional photons to high energies. 
The characteristic energy-loss time due to IC scattering depends on the photon density but is typically longer than the gap light-crossing time, $t_{\mathrm{esc}} \equiv H_\mathrm{gap}/c \sim 10^{-5}$–$10^{-4}\,\mathrm{s}$. 
This limits the instantaneous growth rate of the cascade, as each newly created pair increases the IC scattering rate only modestly.
A key point, however, is that the threshold for producing pair-generating photons is much lower than the maximum attainable energy. 
For our simulation, matching MSPs, the required Lorentz factor is $\gamma_{\mathrm{thr}} \sim 10^{3} \ll \gamma_\mathrm{gap}$, corresponding to photon energies $\varepsilon_\mathrm{thr} \sim 10^{3}$ for two-photon pair production.
For the cooler WD photon field, the corresponding thresholds are higher, $\gamma_{\mathrm{thr}} \sim \varepsilon_\mathrm{thr} \sim 10^{4}$, while remaining well below $\gamma_\mathrm{gap}$ and yielding qualitatively similar behavior.
Consequently, particles can already produce pair-creating photons well before reaching $\gamma_\mathrm{gap}$, enabling early screening of the electric field.

The onset of the cascade is illustrated in Figure~\ref{fig:gap_sim} for the first discharge cycle. 
Starting from a low-energy plasma, the gap is rapidly populated by high-energy photons and pairs across a broad range of altitudes. 
Pair production is most efficient in regions where particles reach $\gamma \sim \gamma_{\mathrm{thr}}$ and photons with $\varepsilon \sim \varepsilon_{\mathrm{thr}}$ quickly create pairs. 
The newly created charges are accelerated by $E_x$, with electrons streaming outward and positrons returning toward the surface (for $\alpha > 0$). 
This plasma generates a current density $j_\pm = m_\pm \eta_{\mathrm{co}} |\beta_s| c$, where $m_\pm = |\eta_\pm / \eta_{\mathrm{co}}|$ is the pair multiplicity, $\eta_\pm = 2 n_s e$, $n_s$ is the number density, and $\beta_s$ is the bulk velocity of species $s$. 
As $|\vec{j}| \gg |\vec{j}_m|$, the electric field is rapidly screened, and Equation~\eqref{eq:maxwell4} reduces to $\partial_t E_x \approx -4\pi j_\pm$.

Once the gap is screened, particle acceleration weakens and pair creation subsides (Figure~\ref{fig:gap_sim}, third panel). 
As the plasma escapes, $\vec{j} \rightarrow 0$, allowing the electric field to rebuild via $\partial_t E_x \approx 4\pi j_m$. 
This leads to a limit-cycle behavior with intermittent pair discharges, analogous to that found in pulsars with stronger magnetic fields \citep{timokhin2010}.
In our simulations, the inner gap remains screened for extended durations, as newly generated pairs replenish the region faster than it is depleted. 
However, this feature is likely overestimated in our 1D setup, where the photon density does not decrease with altitude or diffuse sideways perpendicular to $\vec{B}$. 
Multi-dimensional effects should be explored in future work to better model the pair discharges over many cycles.

\begin{figure}[t!]
\centering
\resizebox{\hsize}{!}{\includegraphics[width=\textwidth]{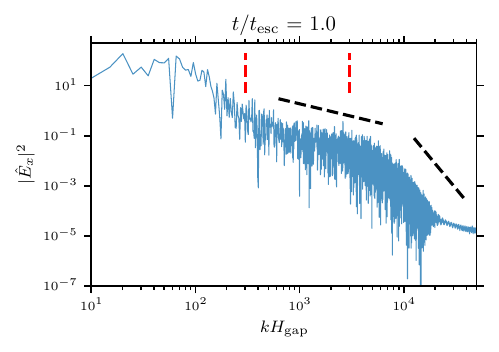}}
\caption{\label{fig:fourier_spectrum} 
Fourier spectra of the electric field for $|\hat{E}_x|^2$.
The spectrum can be approximated as $|\hat{E}_x|^2 \propto k^{-1}$ and $|\hat{E}_x|^2 \propto k^{-5}$ (shown with tilted dashed lines).
The wavenumber range matching the characteristic oscillation frequencies $(\omega_\mathrm{osc}/c)H_\mathrm{gap}$ is shown by the red vertical dashed lines based on the variation in $m_\pm $ and $\langle\gamma^{-3}\rangle$ around this simulation time step.
}
\end{figure}

\section{Coherent Radio Emission} 

The rapid screening of the gap excites electric field oscillations, $\delta E \propto \sin(\omega_\mathrm{osc} t)$, within the discharge region \citep{tolman2022,okawa2024}. 
The characteristic frequency is $\omega_\mathrm{osc} \approx (m_\pm \langle\gamma^{-3}\rangle \omega_B \beta_\mathrm{rot}/t_\mathrm{esc})^{1/2}$, where $\omega_B \equiv e B_0 / m_e c$ is the gyrofrequency \citep{Nattila2026}.
In multidimensional cases, such oscillations can couple to propagating electromagnetic modes \citep{philippov2020}, in particular the superluminal O-mode, which can escape the magnetosphere and power coherent radio emission. 
In our simulations, we find $m_\pm \sim 10$–$100$ and $\langle \gamma^{-3} \rangle \sim 10^{-2}$–$10^{-1}$ during this phase. 
These values imply characteristic emission frequencies $\nu_\mathrm{osc} \sim 0.1$–$1\,\mathrm{GHz}$ for MSPs and $\sim 0.1$–$1\,\mathrm{MHz}$ for WDs.
The estimated characteristic frequencies span broad ranges that overlap the radio bands in which MSPs and WD candidates are observed \citep{lorimer2008,jankowski2018,marsh2016,hurleywalker2022}.
This supports gap discharges as a potential source of coherent electromagnetic fluctuations in both classes of objects.

The power spectra of the simulated oscillations are shown in Figure~\ref{fig:fourier_spectrum}. 
The O-mode-like component (dominated by fluctuations in $E_x$) exhibits a broadband spectrum that can be approximated by a broken power law, $|E_x|^2 \propto k^{-s}$, with slopes $s_1 \approx 1$ and $s_2 \approx 5$.
The high-wavenumber part of the spectrum is, however, time dependent: at $kH_\mathrm{gap} \gtrsim 10^{4}$, the power becomes increasingly suppressed as the oscillation evolves.
The low-$k$ slope is qualitatively comparable to the spectral indices observed in pulsar radio emission, $S_\nu \propto \nu^{-\alpha}$ with $\alpha \sim 1$–$2$ \citep{lorimer2008,jankowski2018}.

\begin{figure}[!t]
\centering
\resizebox{\hsize}{!}{\includegraphics[width=\textwidth,
    trim=0 0 0 1cm,
    clip]{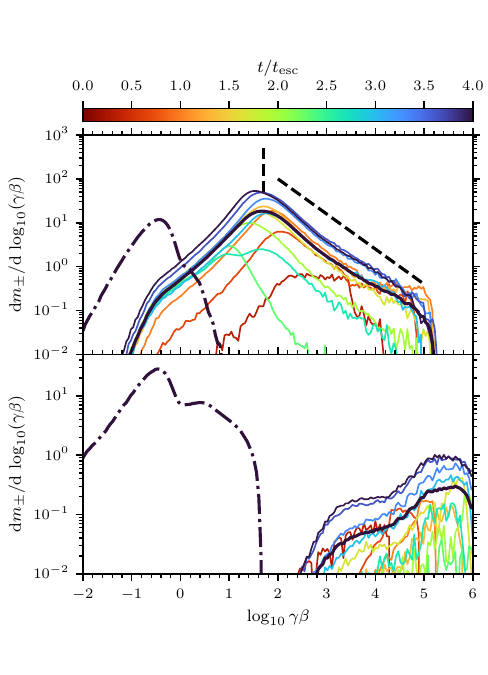}}
\vspace{-1.25cm}
\caption{\label{fig:energy_spectra} 
Distribution of particle momenta for inward moving particles close to the star surface during a discharge cycle for two-photon (top panel) and one-photon (bottom panel) pair creation cases, including time-evolving spectra (thin curves) and average spectrum (thick black curve).
The outgoing electron spectrum in the surface-adjacent cell is also shown, providing a proxy for the thermally injected population (thick dash-dotted curve).
The vertical dashed line in the top panel corresponds to $\gamma = 50$ and the inclined dashed line to a slope of $dm_{\pm}/d\gamma\propto \gamma^{-1.8}$.
}
\end{figure}

\section{Return-Particle Heating}

The momenta of gap-region particles and the near-surface inflowing component are shown in Figure~\ref{fig:energy_spectra} (upper panel).
The return-current distribution peaks at $\gamma \sim 50$ and exhibits a power-law tail $d m_{\pm}/d\ln(\gamma) \propto \gamma^{-0.8}$ extending to $\gamma \sim 10^{5}$.  
Integrating the spectrum yields a pair multiplicity $m_{\pm} \sim 10$ and mean Lorentz factor $\langle\gamma\rangle \sim 600$.   
The inflowing plasma deposits its energy at the polar cap, heating the surface to \citep{salmi2020}
\begin{equation}\label{eq:temperature_estimate}
T \simeq \left(\frac{m_{\pm}\, n_{\mathrm{co}}\, \langle\gamma\rangle\, m_e c^3}{\sigma_{\mathrm{SB}}}\right)^{1/4}.
\end{equation}
Here $n_{\mathrm{co}} \sim 10^{6} (B_{0}/10^{8}\,\mathrm{G}) P^{-1}\,\mathrm{cm}^{-3}$ \citep{Nattila2026}, corresponding to $n_{\mathrm{co}} \sim 10^{9}\,\mathrm{cm}^{-3}$ for MSPs and $n_{\mathrm{co}} \sim 10^{3}\,\mathrm{cm}^{-3}$ for WDs. 
This yields $T \sim 10^{5}\,\mathrm{K}$ ($k_\mathrm{B}T \sim 10^{-2}\,\mathrm{keV}$) for MSPs and $T \sim 10^{4}\,\mathrm{K}$ ($k_\mathrm{B}T \sim 10^{-3}\,\mathrm{keV}$) for WDs, somewhat below the assumed input temperatures, likely reflecting the simplified treatment of energy deposition and radiative transfer.

The return-current energies obtained here are significantly lower than those predicted for regular radio pulsars \citep{timokhin2013,Nattila2026}.  
This difference arises because two-photon pair production—expected to dominate in many MSPs and sufficiently hot WDs—operates at lower photon energies than the one-photon channel.  
Previous studies \citep{harding2011,Harding2022}, based on one-photon pair creation, inferred substantially higher return-current energies, as weaker magnetic fields require higher photon energies to trigger the process (consistent with our one-photon runs; see below and the bottom panel in Figure \ref{fig:energy_spectra}).

The simulated lower-energy return current may affect the neutron-star atmosphere by preferentially heating the upper layers, where the slowest particles deposit their energy \citep{baubock2019,salmi2020}.  
The resulting particle distribution is broadly consistent with the model of \citet{salmi2020} with $\gamma_{\mathrm{min}}=20$ and power-law slope $\delta=2$, which produces a modest high-energy tail and angular emission patterns differing by $10$–$30\%$ from standard deep-heating models.  
Incorporating such atmospheres in NICER analyses of PSR~J0030$+$0451 and PSR~J0740$+$6620 did not significantly alter the inferred radii \citep{salmi2023}, although the effect may be more pronounced for hotter MSPs such as PSR~J0437$-$4715 \citep{choudhury2024}, which also exhibits a pulsed high-energy power-law component \citep{guillot2016,miller2026}.

\section{One-photon pair creation case}

One-photon pair creation can surpass the two-photon process in MSPs for $k_\mathrm{B}T_\mathrm{BB}\lesssim 0.1$ keV (see Appendix \ref{app:estimates}).
We therefore simulate a scenario including Compton up-scattering and one-photon pair creation as the only QED processes.
The parameters are scaled such that $l_\mathrm{mfp}/H_\mathrm{gap} \sim 10^{-2}$ (i.e., close to the estimated value), where $l_\mathrm{mfp}$ is the mean free path for the one-photon process, while maintaining $\gamma_\mathrm{gap} \sim 10^{6}$.

In contrast to the two-photon case (Figure~\ref{fig:gap_sim}), the pair cascade is triggered only near the outer edge of the gap, where particles approach $\gamma_\mathrm{gap}$. 
The newly created pairs partially screen the electric field until they escape, leading again to a limit-cycle behavior with intermittent discharges. 
However, the screening remains incomplete due to the lower pair multiplicity.
The Fourier spectrum of the $E_x$ oscillations exhibits enhanced power at frequencies similar to those in Figure~\ref{fig:fourier_spectrum}, albeit with reduced amplitude. 
However, the predicted oscillation frequency is significantly lower: with $m_\pm \sim 1$ and $\langle \gamma^{-3} \rangle \sim 10^{-12}$--$10^{-10}$, the scaling relation implies a reduction by $\sim 5$ orders of magnitude.
Finally, the return-current particles reach substantially higher energies, $\langle \gamma \rangle \sim 10^{5}$ (Figure~\ref{fig:energy_spectra}), since only the highest-energy photons can produce pairs in the one-photon channel.
This highlights a qualitative difference between the two regimes: two-photon cascades produce dense, low-energy plasmas that efficiently screen the gap, whereas one-photon cascades yield sparse, high-energy return currents with weaker screening.
The estimated surface temperature (based on Equation \ref{eq:temperature_estimate}) is now higher due to larger particle energies but not more than by a factor of $\lesssim 10$.

\section{Discussion}

We demonstrate that pair discharges can be sustained in low-magnetic-field stellar polar caps without curvature radiation, under conditions relevant to MSPs and sufficiently hot WDs.
In this regime, Compton up-scattering can be followed by two-photon pair creation, producing substantially lower-energy return-current particles than previously predicted. 
This has direct implications for MSP atmosphere modeling and, consequently, for X-ray pulse-profile analyses used to constrain the NS equation of state.

The primary limitation of this work is the one-dimensional geometry. 
While the simulations capture the essential microphysics of gap discharges, multidimensional effects are required to determine the spatial structure of the cascade and the resulting surface heating patterns. 
Nevertheless, the 1D approach remains valuable, as it establishes that these QED processes alone can sustain pair cascades under MSP and sufficiently hot WD conditions. 
A second uncertainty arises from the interplay of different QED processes. 
Here we consider one- and two-photon processes separately, whereas in realistic MSPs both may operate simultaneously, depending on surface temperature and magnetic-field geometry. 
In such cases, the return-current spectrum is expected to combine the two regimes shown in Figure~\ref{fig:energy_spectra}.
In addition, we ignored curvature radiation which could become relevant for some MSPs.

Our results are broadly consistent with earlier theoretical estimates that two-photon pair production becomes important for $T_{\mathrm{BB}} \gtrsim 10^6\,\mathrm{K}$ \citep{zhang1998,Harding2002,voisin2018,jones2021}. 
Compton scattering has also been considered in pulsar cascades \citep{luo1996,harding2002ICS}, although previous studies found curvature radiation to dominate in higher-field pulsars \citep{timokhin2015}. 
In contrast, our results show that in low-field systems Compton-driven cascades alone can sustain pair discharges, without invoking strongly multipolar surface fields or extremely small curvature radii \citep{Harding2022,ye2025}. 
This brings MSPs closer to the regime explored in black-hole gap models, where similar QED processes play a central role \citep{levinson2018,yuan2025}. 

Our results also suggest that analogous mechanisms could operate in LPTs if these sources are powered by sufficiently hot, magnetized WDs, although other origins have also been proposed \citep{qu_zhang_2026}.
Their long spin periods and large radio luminosities are difficult to reconcile with standard isolated-NS pulsar models, but WD-powered scenarios can in some cases provide spin-down limits exceeding the inferred radio luminosities (see Figure~4 of \citealt{rea2024}), alleviating the energy-budget problem discussed for isolated NS interpretations of LPTs \citep{rea2026}. 
A WD-pulsar interpretation of LPTs was already proposed by \citet{zhang2005}, although in that model one-photon pair creation required temporary enhancements of the surface magnetic field. 
The mechanism considered here offers a different way to evade the WD death valley, since two-photon pair creation can operate at lower magnetic fields, provided that a sufficiently intense soft-photon field is present.

Observationally, MSPs exhibit relatively stable radio emission, with weak or absent sub-pulse drifting, nulling, and mode changing \citep{parthasarathy2021}. 
In standard interpretations, sub-pulse variability is usually associated with spatially nonuniform and time-dependent discharge regions \citep{ruderman1975}. 
Our simulations indicate that the relative stability of MSP radio emission may instead reflect the cascade microphysics: in the MSP regime, pair production proceeds more continuously, with weaker screening--unscreening cycles because the background photon bath enabling the pair creation is more diffuse and the cascade region more extended. 
Such an extended discharge region can more easily average over local fluctuations and suppress observable sub-pulse variations. 
In contrast, this mechanism cannot explain the highly sporadic radio activity of LPTs \citep{horvath2026}, where additional global magnetospheric effects are likely required to account for the observed variability. 
If LPTs are powered, for example, by a binary system containing a WD, the cascade may experience a strongly varying $\alpha$ due to strongly evolving magnetospheric twist and hence a time-dependent $j_m$.

In summary, we identify a new regime of pair cascades driven by Compton scattering and two-photon pair production, applicable to some MSPs and WDs. 
These cascades produce softer return currents, modify polar-cap heating, and naturally generate coherent radio emission. 
Our results highlight the importance of radiative processes beyond curvature emission and provide a direct link between kinetic gap physics and observable radio emission.

\begin{acknowledgments}
This work is supported by ERC grant (ILLUMINATOR, 101114623) and Research Council of Finland, the Centre of Excellence in Neutron-Star Physics (project 374063). 
The views and opinions expressed are, however, those of the authors only and do not necessarily reflect those of the European Union or the European Research Council. 
Neither the European Union nor the granting authority can be held responsible for them. 
T.S. acknowledges funding by the Research Council of Finland grant No.~368807.
T.S. and J.N. acknowledge useful discussions with A. Timokhin, J. Benáček, E. van Woerkom, and Y. Qu.
The authors thank the Finnish Computing Competence Infrastructure (FCCI) for supporting this project with computational and data storage resources and for maintaining the HILE cluster.
The authors used OpenAI GPT-5.6 models for assistance with equation checking and manuscript editing.
\end{acknowledgments}

\newpage

\appendix

\section{Analytical Estimates}\label{app:estimates}

\begin{figure*}[t!]
\centering
\includegraphics[clip, trim=0.0cm 0.0cm 0.0cm 0.0cm, width=0.45\textwidth]{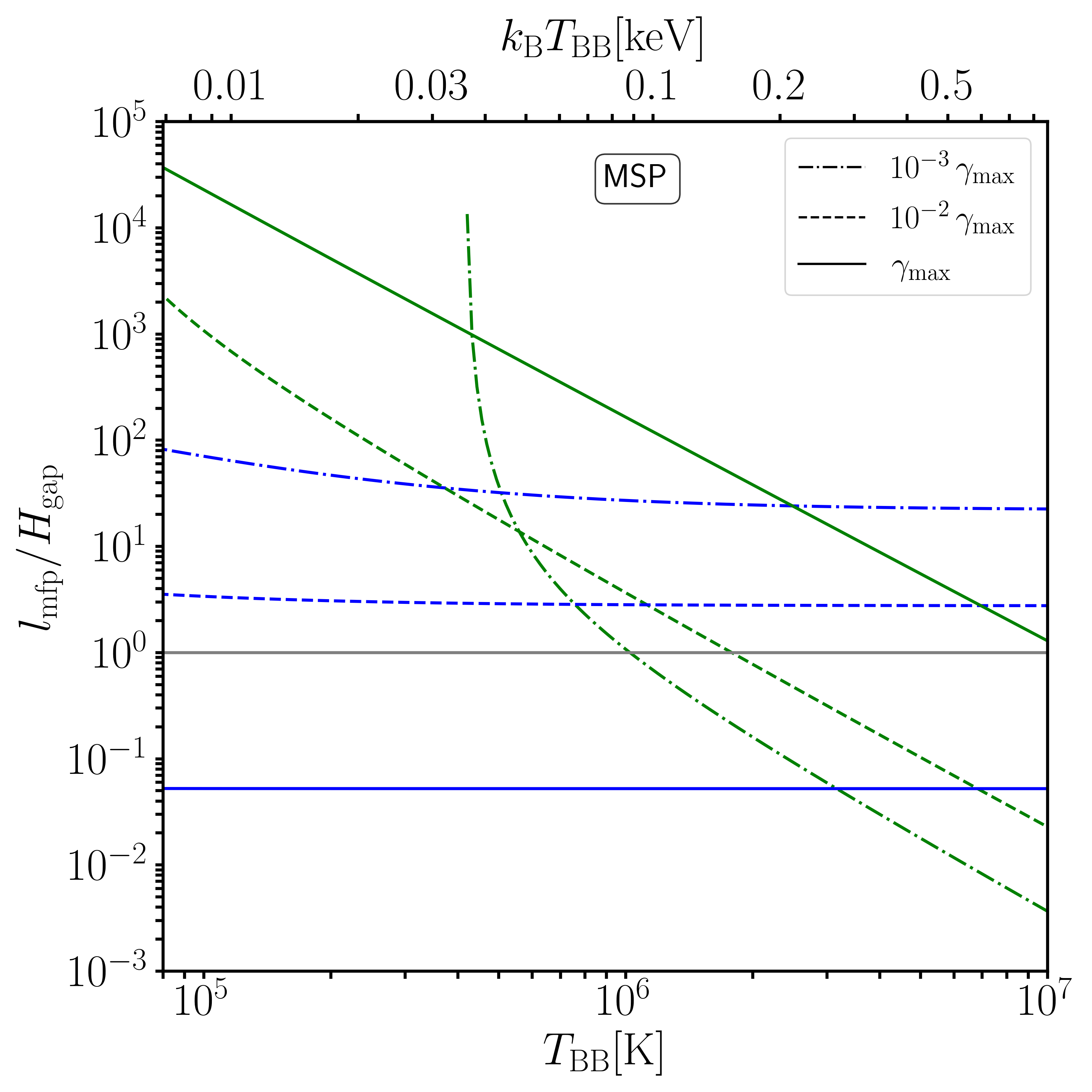}
\includegraphics[clip, trim=0.0cm 0.0cm 0.0cm 0.0cm, width=0.45\textwidth]{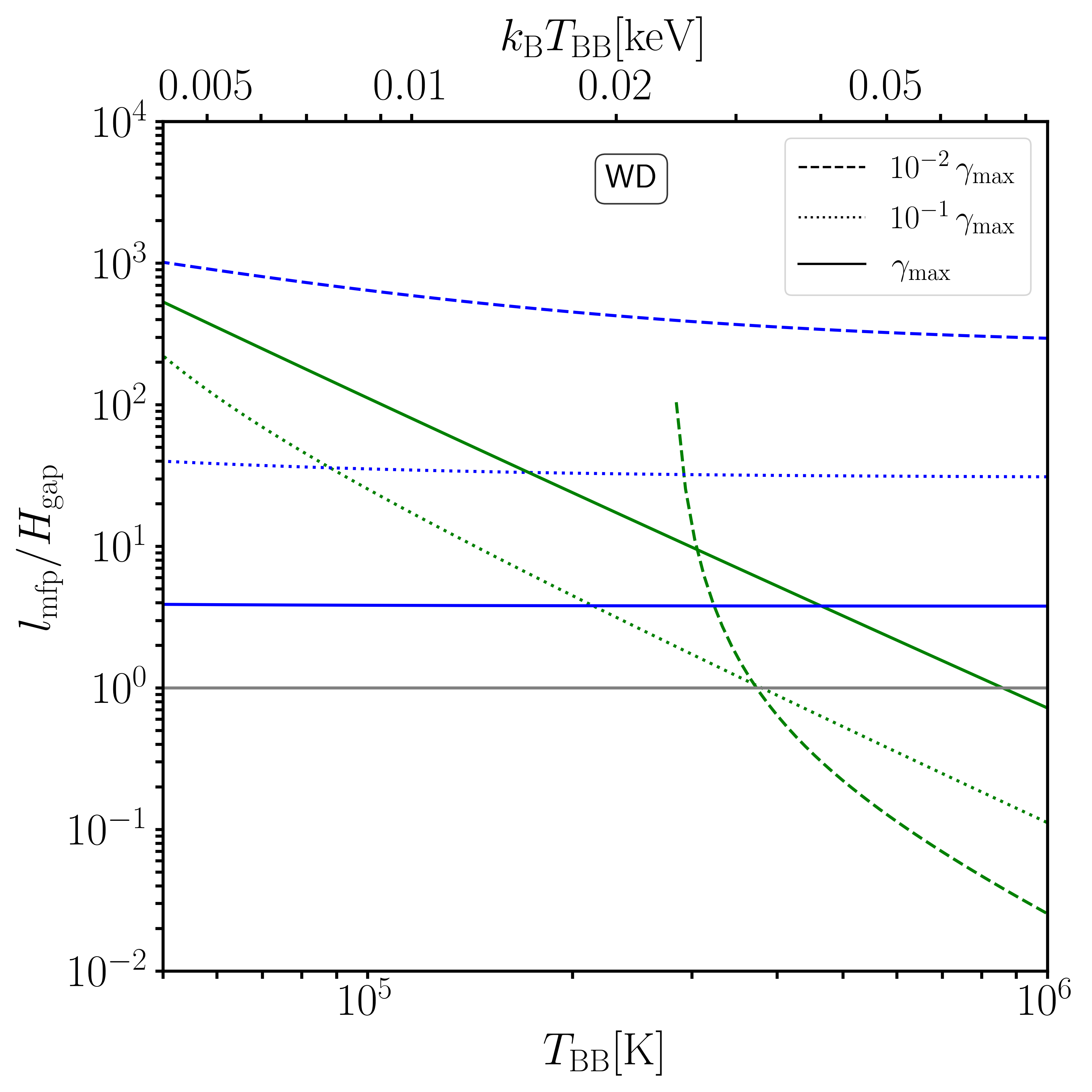}
\caption{\label{fig:lmfps} 
Mean free paths of one-photon (blue) and two-photon (green) pair creation processes, normalized by the polar-gap height, as function of temperature of thermal photons $T_\mathrm{BB}$ in case of our MSP (left panel) and WD (right panel) setups.
The polar gap size is estimated as $H_\mathrm{gap} = R_\mathrm{pc}$, given in Equation \ref{eq:polar_cap_size}.
The mean free paths are calculated using Equations \ref{eq:lmfp_pair_1phot} and \ref{eq:lmfp_pair_2phot}, assuming the high-energy photon is produced by scattering once from a particle with different Lorentz factors (presented with different curve styles).
$\l_\mathrm{mfp}/H_\mathrm{gap} = 1$ is shown with a gray line.
}
\end{figure*}

\subsection{Polar Cap Estimates}
Here we present useful polar-cap quantities. 
The polar cap size $R_\mathrm{pc}$ can be estimated using the separatrix angle between the last closed field line and the dipole axis $\sin\theta_\mathrm{pc} \approx \sqrt{ R_\star/R_\mathrm{LC}}$, where $R_\star$ is the stellar radius, $R_\mathrm{LC} = cP/(2\pi)$ is the light cylinder radius, and $P$ is the spin period of the star.
The polar-cap radius is therefore 
\begin{equation}\label{eq:polar_cap_size}
    \frac{R_\mathrm{pc}}{R_\star} \approx \sin\theta_\mathrm{pc} \approx \Big(\frac{2\pi}{c}\Big)^{1/2}R_\star^{1/2}P^{-1/2}\, .
\end{equation}
For MSPs $R_\mathrm{pc}/R_\star \approx 0.3$, assuming $R_\star = 10^6$ cm and $P=2$ ms.
For WDs $R_\mathrm{pc}/R_\star \approx 0.01$, assuming $R_\star = 10^9$ cm and $P=10^{3}$ s. 

The velocity at distance $R_\mathrm{pc}$ from the spin axis is then
\begin{equation}
    \beta_\mathrm{rot} \approx \frac{2 \pi R_\mathrm{pc}}{cP}  \approx  \Big(\frac{2\pi}{c}\Big)^{3/2}R_\star^{3/2}P^{-3/2}\,.
\end{equation}
This yields $\beta_\mathrm{rot} \approx 3 \times 10^{-2}$ for the MSP parameters and $\beta_\mathrm{rot} \approx 3 \times 10^{-6}$ for the WD case.

\subsection{Particle Dynamics in the Gap}
The charged particle dynamics is governed by
\begin{equation}
    \frac{\ud \vec{u} }{\ud t} = \frac{q}{m_e} \left( \vec{E} + \frac{\vec{v}}{c} \times \vec{B} \right) - \frac{\vec{F}_\mathrm{rad}}{m_e} \, ,
\end{equation}
where $\vec{u}=\gamma\vec{v}$ is the proper velocity. In the strongly magnetized limit, the motion reduces to the one-dimensional beads-on-wire approximation with the spatial coordinate $s$ along $\vec{B}$. 
For relativistic upward motion, $u \equiv \beta\gamma c \rightarrow \gamma c$ and $P_\mathrm{rad} \equiv \vec{F}_\mathrm{rad}\cdot\vec{\beta}c \rightarrow F_\mathrm{rad}c$. 
The evolution equation becomes
\begin{equation}
    \frac{\ud \gamma}{\ud t} \simeq \frac{q E_\parallel}{m_e c} - \frac{P_\mathrm{rad}}{m_e c^2} \, .
\end{equation}

The parallel electric field near the polar cap is approximated as \citep[e.g.,][]{daugherty1982,timokhin2010}
\begin{equation}
E_\parallel = 
\begin{cases}
    -\frac{\Delta V}{H_{\rm gap}}, \quad & s \leq H_\mathrm{gap} \\
    0,                  \quad & s > H_\mathrm{gap}
\end{cases}
\end{equation}
where $H_{\rm gap}$ is the gap height and $s=0$ at the stellar surface. 
We adopt the commonly used estimate $H_{\rm gap} \sim R_\mathrm{pc}$.

The maximum Lorentz factor reached across the full potential drop (neglecting radiation losses) is
\begin{equation}\label{eq:gam_gap}
\begin{split}
    \gamma_{\mathrm{gap}} 
    = \int_{0}^{H} \frac{\ud \gamma }{\ud s} \ud s
    = \frac{e \Delta V}{m_e c^2} 
    \sim \frac{e B_{0} \beta_\mathrm{rot} R_\mathrm{pc}}{2m_e c^2} \\
    \sim  \frac{2 \pi^2 e}{m_e c^4}\Big(\frac{B_{0}R_\star^{3}}{P^{2}}\Big) \, .
\end{split}
\end{equation}
For the fiducial MSP parameters we obtain $\gamma_\mathrm{gap}\sim10^{8}$, while for WDs $\gamma_\mathrm{gap}\sim10^{6}$, assuming $B_0=10^{8}$ G.

In principle, radiative losses from curvature radiation or Compton scattering may prevent particles from reaching the maximum Lorentz factor. 
Assuming curvature losses, one obtains $\gamma_\mathrm{rad}\sim10^{7}$ for a radiative balance \citep[see][for equations]{Nattila2026}, both in MSPs and WDs (assuming a curvature radius $R_\mathrm{curv}\sim R_\star$). 
This value is comparable to $\gamma_\mathrm{gap}$ and therefore does not significantly limit the particle acceleration, at least in WDs. 

The drag caused by Compton losses can be obtained from 
\begin{equation}\label{eq:P_compt}
    \frac{\ud \gamma}{\ud t} \Bigg|_\mathrm{comp}
    = - \frac{P_\mathrm{C}}{m_e c^2} 
    \approx - \frac{4}{3}\frac{\sigma_\mathrm{T}c\gamma^{2}U_{\rm BB}}{m_e c^{2}}f_{\rm KN}(\gamma \varepsilon_{\rm BB}),
\end{equation}
where $f_{\rm KN}(\gamma\varepsilon_{\rm BB})$ is the Klein-Nishina correction factor and $\varepsilon_{\rm BB}$ is the characteristic energy of thermal background photons.
The photon energy density is $U_{\rm BB} = \varepsilon_{\rm BB} n_{\rm BB} \approx 2.7 k_{\rm B} T_{\rm BB} n_{\rm BB}$, and the photon number density is
\begin{equation}\label{eq:nx}
    n_{\rm BB} \approx 1.202 \times 16 \pi \left( \frac{k_{\rm B}T_{\rm BB}}{hc}\right)^{3}.
\end{equation}
In a steady state ($\ud\gamma/\ud t=0$), acceleration should balance radiative drag, yielding
\begin{equation}\label{eq:gam_rad}
\begin{split}
  \gamma_{\mathrm{rad}}^{2}f_{\rm KN}(\gamma_{\rm rad}\varepsilon_{\rm BB})  = 0.28 \frac{m_e c^2}{k_{\rm B}T_{\rm BB}}\frac{\gamma_\mathrm{gap}}{\sigma_\mathrm{T}n_{\rm BB} R_{\rm pc}}.
\end{split}
\end{equation}
However, solving this equation numerically for MSP parameters with $k_BT_{\rm BB} < 1$ keV and WD parameters with $k_BT_{\rm BB} < 0.1$ keV, we find that no solutions with $\gamma_{\rm rad}<\gamma_{\rm gap}$ exist due to the strong Klein–Nishina suppression of the Compton scattering cross section \citep[see Appendix C of][for an exact expression of $f_{\rm KN}$]{moderski2005}. 
Radiative Compton drag therefore cannot balance electric acceleration, and particles can reach the full gap potential or the curvature equilibrium. 
Compton scattering is nevertheless important when producing high-energy photons capable of pair creation, as discussed next.

\subsection{Production of High-energy Photons}
In high-field radio pulsars, curvature radiation is expected to be sufficiently energetic to produce photons capable of pair creation. 
However, in MSPs and WDs the characteristic curvature photon energies, $\varepsilon_\mathrm{curv} \sim 10^{-1}$–$10^{4}$ \citep{Nattila2026}, are insufficient for efficient one-photon pair production (see below). 
In hotter MSPs, two-photon pair creation from curvature photons of the most energetic particles may still operate (see below), but for simplicity we neglect this channel.

In low-field MSPs and WDs, inverse Compton up-scattering provides an important alternative mechanism for producing high-energy photons. 
In a single scattering, the photon energy increases approximately as \citep{rybicki1979,aharonian2004}
\begin{equation}\label{eq:energy_gain_in_compton}
\varepsilon_{\mathrm{IC}} \approx \frac{4\gamma^{2}\varepsilon_{\mathrm{BB}}}{1 + 4 \gamma \varepsilon_{\mathrm{BB}}},
\end{equation}
for head-on collisions. 
This expression remains valid into the Klein–Nishina regime ($\gamma \varepsilon_{\mathrm{BB}} \gtrsim 1$), which is relevant for most of our parameter space. 
In this regime, a single scattering can remove a significant fraction of the particle energy.

The characteristic cooling length of an individual particle is
\begin{equation}\label{eq:lmfp_compton}
l_{\mathrm{mfp}} \approx ct_{\mathrm{comp}} \approx \frac{\gamma m_{e}c^{3}}{P_\mathrm{C}} \approx \frac{3}{4}\frac{m_{e}c^{2}}{\gamma\sigma_\mathrm{T}U_\mathrm{BB}f_{\rm KN}(\gamma \varepsilon_{\rm BB})}.
\end{equation}
For MSP parameters with $k_BT_{\rm BB}\lesssim1$ keV and WD parameters with $k_BT_{\rm BB}\lesssim0.1$ keV, we find $l_\mathrm{mfp} > H_\mathrm{gap}$ for particles with $\gamma \sim \gamma_\mathrm{max} \equiv \min(\gamma_\mathrm{gap},\gamma_\mathrm{rad})$. 
This implies that individual high-energy particles typically travel distances comparable to or larger than the gap size before scattering.
However, particles at lower Lorentz factors experience weaker Klein–Nishina suppression and therefore scatter more efficiently. 
Combined with a particle number density $n_{\rm co} \sim 10^{6} (B_{0}/10^{8}\,\mathrm{G}) P^{-1}\,\mathrm{cm}^{-3}$ \citep{Nattila2026}, this ensures that inverse Compton scatterings still occur within the gap.

\subsection{One-photon Pair Creation}
The one-photon pair creation process in a magnetic field ($\gamma + [B] \rightarrow e^+ + e^-$) has a mean free path given by \citep{Nattila2026}: $l_\mathrm{mfp} \approx R_{\mathrm{curv}} \chi_{\varepsilon}^{\rm a}/(\varepsilon b)$, where $b = B_{0}/B_\mathrm{Q}$ and $\chi_{\varepsilon}^{\rm a}$ is the photon quantum parameter at which the optical depth satisfies
\begin{align} \label{eq:lmfp_pair_1phot}
\begin{split}
\tau_{\rm BW}(\chi_{\varepsilon}^{\mathrm{a}}) =0.086 \frac{\alpha_\mathrm{f} R_{\mathrm{curv}}}{\lamC \varepsilon^{2} b} (\chi_{\varepsilon}^{\mathrm{a}})^{3} e^{-\frac{8}{3\chi_{\varepsilon}^{\mathrm{a}}}}=1.
\end{split}
\end{align}
We solve this equation numerically to estimate $\chi_{\varepsilon}^{\rm a}$ and the corresponding mean free paths for photons that have undergone a single Compton up-scattering by particles with Lorentz factors up to $\gamma_\mathrm{max}$ (blue curves in Figure~\ref{fig:lmfps}).
We find that one-photon pair creation becomes efficient ($l_\mathrm{mfp} < H_\mathrm{gap}$) for the highest particle energies in MSPs, but not in WDs. 
A lower temperature increases slightly the mean free path because the up-scattered photon energies are reduced when $\gamma \ll \gamma_\mathrm{max}$.
Compton scattering is essential for this channel, since even maximally energetic curvature photons yield $l_\mathrm{mfp}/H_\mathrm{gap} \sim 10$ in MSPs and $l_\mathrm{mfp}/H_\mathrm{gap} \sim 10^{7}$ in WDs.

\subsection{Two-photon Pair Creation}
For two-photon pair creation ($\gamma + \gamma \rightarrow e^+ + e^-$), the mean free path can be estimated as \citep[see Equation~4.7 in][]{coppi1990}
\begin{align} \label{eq:lmfp_pair_2phot}
\begin{split}
l_\mathrm{mfp} \approx \frac{\varepsilon_{12}^{3}}{0.652 \sigma_{\mathrm{T}}(\varepsilon_{12}^2-1)\ln(\varepsilon_{12})H(\varepsilon_{12}-1)n_{\mathrm{BB}}},
\end{split}
\end{align}
where $\varepsilon_{12}=\varepsilon_1\varepsilon_2$, and $H(\varepsilon_{12}-1)$ is the Heaviside function enforcing the threshold condition $\varepsilon_{12}>1$. 
Here $\varepsilon_1$ and $\varepsilon_2$ are the photon energies.
The resulting mean free paths are shown in Figure~\ref{fig:lmfps}, assuming that one photon has the characteristic thermal energy $\varepsilon_{\rm BB}$ and the other is produced via a single Compton up-scattering by particles with Lorentz factors up to $\gamma_\mathrm{max}$ (green curves).
We find that two-photon pair creation can proceed efficiently in both MSPs and WDs at sufficiently high $T_{\rm BB}$ when the particle Lorentz factor is a fraction of $\gamma_\mathrm{max}$. 
For smaller $\gamma$ and $T_{\rm BB}$, the up-scattered photon energy falls below the pair-production threshold, $\varepsilon_{12}<1$, and the mean free path formally diverges.
In MSPs, curvature photons can also undergo two-photon pair creation when interacting with thermal photons (yielding $l_\mathrm{mfp}/H_\mathrm{gap}\sim0.1$ at $k_{\rm B}T_{\rm BB}=0.4$ keV), whereas in WDs the corresponding photon energies remain below the threshold.

\section{Supplementary Methods}\label{app:supplementary}

Here we provide additional details on the numerical parameters and simulation techniques, complementing the description in Section~\ref{sect:setup}. 
The simulation code is available at \url{https://www.github.com/hel-astro-lab/runko} \citep{runkoDOI}, where we used the commit \texttt{86946cc} from the archived branch \texttt{v4-qed}. 
The simulation and data-analysis scripts are available at \url{https://github.com/hel-astro-lab/polarcaps}.

The simulations use a second-order finite-difference field solver to advance the electromagnetic fields and a relativistic Boris pusher to evolve the particle momenta. 
The current is deposited using the charge-conserving ZigZag algorithm. 
In addition to the particle current, we include a prescribed virtual current that imposes the externally driven twist, controlled by the parameter $\alpha$ introduced in Section~\ref{sect:setup}. 
The time step satisfies the Courant--Friedrichs--Lewy (CFL) stability condition with a Courant factor of $0.45$.

The QED interactions are treated using the same module and adaptive Monte Carlo routines as described in \citet[][Supplementary Material]{nattila2024}. 
For numerical feasibility, the QED interaction rates were rescaled such that each interacting computational particle represents $\mathcal{N}_\mathrm{mp} = 10^{8}$ physical particles. 
Equivalently, this corresponds to modifying the effective electron radius entering the interaction probabilities; see Equation~(A.18) of \citet{runko}. 
The injected number of thermal photons was also reduced, and scaled values of the fiducial magnetic-field strength and surface rotation velocity were used in some simulations. 
These rescalings were chosen to make the simulations computationally tractable while preserving the target mean free paths of the included QED processes; see Appendix \ref{app:estimates}. 
Finally, QED interactions were disabled in the region $x > H_\mathrm{gap}$, providing a simple prescription for a finite active gap length.

We also performed simple numerical convergence tests for the two-photon pair-creation simulations. 
In one test, the spatial resolution was reduced by a factor of two, while in another the particle loading corresponding to the co-rotation density was reduced from 16 to 4 particles per cell per species. 
Neither modification led to major qualitative or quantitative changes in the discharge evolution or in the resulting particle and photon distributions. 
This indicates that the main conclusions are not sensitive to these numerical choices.

\bibliography{refs}{}
\bibliographystyle{aasjournalv7}

\end{document}